\documentclass[RNAAS]{aastex62}

\newcommand\kms{km\,s$^{-1}$}	

\begin{document}

\title{A comparison between RAVE DR5 and Gaia DR2 Radial Velocities }

%% Note that the corresponding author command and emails has to come
%% before everything else. Also place all the emails in the \email
%% command instead of using multiple \email calls.
\correspondingauthor{Matthias Steinmetz}
\email{msteinmetz@aip.de}

\author[0000-0001-6516-7459]{Matthias Steinmetz}
\affiliation{Leibniz-Institut f\"ur Astrophysik Potsdam (AIP), An der Sternwarte 16, 14482 Potsdam, Germany}

\author{Toma\v{z} Zwitter}
\affiliation{Faculty of Mathematics and Physics, University of Ljubljana, Jadranska 19, 1000 Ljubljana, Slovenia}

\author{Gal Matijevic}
\affiliation{Leibniz-Institut f\"ur Astrophysik Potsdam (AIP), An der Sternwarte 16, 14482 Potsdam, Germany}

\author{Alessandro Siviero}
\affiliation{Department of Physics and Astronomy, University of Padova, Asiago, Italy}

\author{Ulisse Munari}
\affiliation{INAF National Institute of Astrophysics, Astron. Obs. Padova, Asiago, Italy}

%% Note that RNAAS manuscripts DO NOT have abstracts.
%\facility{UKST}

%% Start the main body of the article. If no sections in the 
%% research note leave the \section call blank to make the title.
\section{} 

The Radial Velocity Experiment (\cite{2006ApJ...132..1645}) is a massive spectroscopic campaign of some 460\,000 stars in the southern hemisphere. RAVE spectra were taken with the 6dF multi-object spectrograph on the 1.2m UK Schmidt telescope of the Australian Astronomical Observatory (formerly Anglo-Australian Observatory). RAVE spectra are taken at a resolution of  $R\approx 7500$ in the IR Calcium triplet region (8410 - 8795\,\AA). Observations for RAVE were taken between April 2003 and April 2013. 

On 25 April 2018, the 2nd data release of the ESA mission Gaia was published (\cite{2018A&A...616A...1G}), featuring radial velocities for some 7 million targets. The Radial Velocity Spectrometer (RVS) of Gaia also operates in the Ca triplet region, though at a somewhat higher resolution of $R=11000$. This research note presents a  comparison of radial velocities between the most recent data release of RAVE (\cite{2017AJ...153..75K}) with Gaia DR2. RAVE DR5 and Gaia DR2 have 450587 stars in common. In particular we would like to shed light on a small subset of joint targets (707 stars) that exhibit a constant velocity offset of $+105$ \kms (see also \cite{arXiv:1807.11716}) or $-76$ \kms, respectively. Since RAVE provided the largest subset of targets for the Gaia pipeline verification, the two data sets are not fully independent.

\begin{enumerate}
\item Figure 1 compares the radial velocities published in Gaia DR2 with those published in RAVE DR5. Overall this comparison confirms the excellent agreement between those two data sets. The velocity differences can well be matched with two Gaussians with FWHM of $1.2$ \kms and $3.6$ \kms, respectively. There is a systematic offset of about $-0.32$ \kms, consistent with the findings published in \cite{arXiv:1807.11716}, who find an offset of $-0.29$ \kms. The offset is also comparable to the offset found between Gaia DR2 and other ground-based spectroscopic surveys in a similar magnitude range, such as APOGEE (\cite{2018A&A...616A...6S}),  indicative that the source for this offset may at least partially be related to the radial velocity zeropoint of Gaia DR2.  The difference is also within the error estimates published in the RAVE data release papers based in internal template matching errors, errors with external samples, and errors derived from a subset of stars with repeat observations. 

\item A further analysis exhibits no systematic tendency with RAVE derived effective temperatures for stars with 4000\,K$<T_{\rm eff}<$7000\,K. Stars cooler than 4000\,K exhibit a somewhat smaller shift of $-0.1$ \kms. For stars hotter than 7000\,K (a small subset of the RAVE sample), the accuracy of the radial velocity deteriorates resulting in a larger systematic shift and a considerably increased FWHM, owing to the increasing dominance of broad Paschen lines at the expense of a less prominent Calcium triplet. With increasing SNR, the prominence of the $1.2$ \kms Gaussian increases, and that of the $3.6$ \kms Gaussian decreases. Dwarfs stars (log $g >3.5$) have a lower fraction of stars in the $1.2$ \kms Gaussian than giant stars. There is a very mild tendency that the velocity shift between RAVE DR5 and Gaia DR2 changes with metallicity. This effect amounts to about $0.5$ \kms between $[Fe/H]<-1$ and  $[Fe/H]>0$.

\item Figure 1 (left) plots the radial velocities published in RAVE DR5 against those in Gaia DR2. This figure shows a small group of 707 stars that exhibit a velocity offset of $\sim 105$ \kms or $\sim -76$ \kms. The first group has independently been identified by \cite{arXiv:1807.11716}. Thanks to the availability of additional data including spectra, observing data and setup, and fiber allocation, we are in a position to identify the cause for this velocity offset. While this group of stars exhibit on average a somewhat lower signal-to-noise than the RAVE DR5 catalog in general, there is no clear correlation of observing date or epoch, stellar parameters, signal-to-noise or any other parameter published in the DR5 catalog, technically or astronomically. However a deeper inspection exhibits that these objects can be found almost exclusively near the edge of the robotically positioned field plate, i.e.\ 
at very low or very high fiber number. The camera used in the 6dF spectrograph has a very fast focal ratio resulting in a large field curvature at the edges of the focal plane. The wavelength calibration, i.e. the mapping between the column number of a CCD pixel to the corresponding wavelength, thus rapidly changes between adjacent fibres near the edges of the focal plane. In a small number of fields our reduction pipeline encountered a problem in propagating the wavelength calibration between neighboring fibres near the edges of the field plate, in particular for observations done just before fibre repairs, i.e. when live fibres were separated by large gaps of broken ones. Such instances are clearly indicated in the corresponding log files which have now been recovered. In total we identified {\bf 707} stars with questionable wavelength calibration. In 82\%\ of these cases the offset between RAVE DR5 and Gaia DR2 radial velocity is in the $+105 \pm 15$~\kms range. These spectra belong to the first 10 fibres of the field plate. Another 3\%\ of the spectra have velocity offset in the $-76 \pm 4$~\kms range and were observed with the fibres 148-150. Even though the majority of these do not exhibit any particularity in the radial velocity, we recommend excluding them from future studies.

\end{enumerate}

We have published a CSV-file with the RAVE\_OBS\_IDs of the suspicious targets on the RAVE DR5 survey website \url{http://www.rave-survey.org}. The corresponding author is also happy to email this file upon request. A more detailed analysis will be published in an upcoming paper featuring the final data release of the RAVE survey (Steinmetz et al., in preparation).

%% An example figure call using \includegraphics
\begin{figure}[h!]
\begin{center}
\includegraphics[scale=0.85,angle=0]{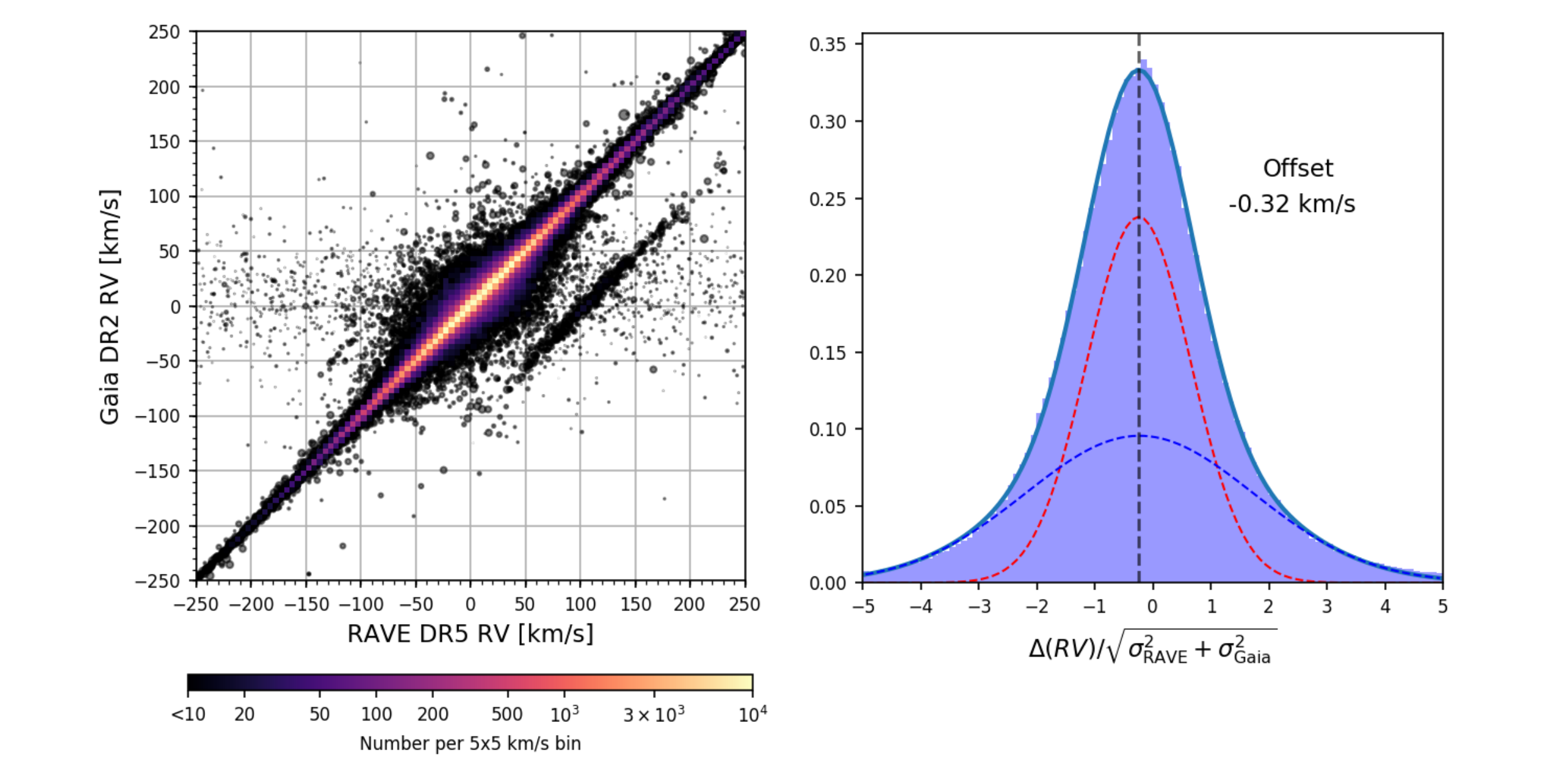}
\caption{Left: Radial Velocity derived from Gaia DR2 vs those from RAVE DR5. Right: Distribution of radial velocity differences between Gaia DR2 and RAVE DR5. The black line compares this distribution function with a fit using two Gaussian with a FWHM of x (red) and y(blue) km/s, respectively. A list with the RAVE\_OBS\_IDs of the outliers is attached. \label{fig:1}}
\end{center}
\end{figure}

%% An example table using AASTeX's deluxetable. Note that since
%% only one figure OR one table is allowed this is commented out.
%\begin{deluxetable}{ccl}
%\tablecaption{Example table some English and Greek letters\label{tab:1}}
%\tablehead{
%\colhead{Index number} & \colhead{English} & \colhead{Greek}
%}
%\startdata
%1 & a & alpha ($\alpha$) \\
%2 & b & beta ($\beta$) \\
%3 & c & gamma ($\gamma$) \\
%4 & d & delta ($\delta$) \\
%5 & e & epsilon ($\epsilon$) \\
%\enddata
%\tablecomments{Long tables should only show a short example with the full
%version as a machine readable table with the article.}
%\end{deluxetable}  

\acknowledgments

\end{document}